\documentstyle{article}

\def\abstract#1{\vskip 7mm 
        \begin{center}{\large Abstract}\par \smallskip
                \begin{minipage}[c]{12cm}
                        \small #1
                \end{minipage}
        \end{center}
}
\def\title#1{\begin{center}{\Large\bf #1}\end{center}}
\def\author#1{\vskip 5mm \begin{center}{#1}\end{center}}
\def\address#1{\begin{center}{\it #1}\end{center}}

\newtheorem{theorem}     {Theorem}
\newtheorem{lemma}       {Lemma}
\newtheorem{proposition} {Proposition}
\newtheorem{definition}  {Definition}

\begin{document}
\title{Global existence problem in 
$T^{3}$-Gowdy symmetric IIB superstring cosmology}

\author{Makoto Narita\footnote{maknar@aei-potsdam.mpg.de}} 

\address{Max-Planck-Institut f\"ur Gravitationsphysik, \\
Albert-Einstein-Institut, 
Am M\"uhlenberg 1, 
D-14476 Golm, \\
Germany
}

\abstract{We show global existence theorems for Gowdy symmetric 
spacetimes with type IIB stringy matter. 
The areal and constant mean curvature 
time coordinates are used. 
Before coming to that, it is shown that 
a wave map describes evolution of this system.}

PACS: $02.30.J_{r}, 04.20.D_{W}, 04.20.E_{X}, 98.80.J_{k}$
\section{Introduction}
\label{intro}

The strong cosmic censorship (SCC) conjecture, which states that 
generic initial data sets have maximal Cauchy developments which 
are locally inextendible as Lorentz manifolds \cite{K,KN},  
is the most important and unsolved problem 
in classical general relativity. 
In order to prove this conjecture two steps are needed. 
The first step is: 
Solve global Cauchy problem of the Einstein-matter equations 
in a suitable time coordinate. 
The second is : 
Show inextendibility of the maximal Cauchy development. 
This article concerns the first step.

The SCC conjecture is entirely out of reach 
at the present time. A reason is that 
the Einstein-matter equations are nonlinear hyperbolic partial 
differential equations (PDEs) in 
some gauge and then, it is very hard to solve global Cauchy problem of them 
without any restrictions and/or simplifications.  
Such restricted and simpler systems can be obtained by looking for 
solutions invariant under a smooth, effective action of isometry groups. 
In this paper we shall consider Gowdy symmetric spacetimes which have 
$U(1)\times U(1)$ isometry group on compact spacelike hypersurfaces. 
The resulting system of PDEs is one of $(1+1)$ semi-linear wave equations.

It is natural to expect that global structure of cosmological 
spacetimes must 
be ultimately found on 
superstring/M-theory
which is the most promising candidate for 
the unified theory including general relativity.  
Then, it is of prime importance to study the possible applications of 
superstring/M-theory to global existence problems.

In our previous work, the Einstein-Maxwell-dilaton-axion (EMDA) system, 
which comes from the low energy effective heterotic superstring theory,
have been considered \cite{N}. 
An important feature of the heterotic theory 
is that all terms of the fields in the effective 
action couple directly to the dilaton field, i.e., 
they consist of Neveu-Schwarz$-$Neveu-Schwarz (NS-NS) sector and 
there is no Ramond$-$Ramond (R-R) sector. 
In the recent research, 
it is suggested that the existence of the R-R fields
influences dynamics of spacetimes \cite{LWC}. 
Therefore, we would like to consider type IIB superstring theory, 
whose all bosonic degree of freedom consist of both NS-NS and R-R sectors. 
This is one motivation to consider the IIB theory.

Another motivation why we consider IIB theory is that the theory has 
a non-perturbative duality, that is, self-S-duality. 
As general assumptions, 
one ignores higher derivative terms (i.e. strong curvature effects) 
in the effective action and/or  
string-loop (i.e. quantum) corrections when 
the global structure of spacetimes is considered 
in superstring/M-theory, since nobody know the {\it complete} theory. 
There is no much sense in including some 
of higher derivative terms at the present time. 
Contrarily, 
string-loop corrections can been considered by using S-duality. 
It has been conjectured that 
the type IIB superstring theory has self-S-duality, 
that is, the IIB theory 
at coupling constant $g_{s}^{({\rm IIB})}$, 
which controls the importance of string-loop corrections, 
is equivalent to the IIB theory 
at coupling constant $1/g_{s}^{({\rm IIB})}$ in ten-dimensions \cite{T}.
This means that the IIB theory is valid in the region of spacetimes 
where quantum corrections for fundamental fields including 
gravity are needed. 
Then, the self-S-duality of the IIB theory 
is appropriate for our purpose that is to analyze spacetimes globally, i. e. 
from the vicinity of the singularity to the future.

In this paper global existence of solutions of the Gowdy symmetric 
spacetimes with type IIB stringy matter will be proved for 
arbitrary large initial data in the areal coordinate.  
The essential ingredients of the proof are 
light cone argument \cite{MV} and 
Christodoulou-Tahvildar-Zadeh's identity \cite{CTZ1}.  
Before coming to that, we need to show that a wave map 
is equivalent to the system of evolution 
equations of the Einstein-matter equations. 
Furthermore, we will obtain results that 
the spacetimes have crushing singularities into the past and 
the mean curvature of the foliation tends uniformly to zero into the future. 
Thus, global existence of constant mean curvature (CMC) foliations 
is shown by established arguments.

\section{Gowdy symmetric spacetimes in IIB superstring theory}
The low-energy limit of the type IIB superstring is given 
by $N=2$, $D=10$ chiral supergravity. 
The massless bosonic fields in the type IIB superstring theory come 
from both NS-NS and R-R sectors. 
The NS-NS sector has the graviton $g_{MN}$, an anti-symmetric tensor field 
$B^{(1)}_{MN}$ and the dilaton field $\Phi$, 
where $M$, $N$ run from $0$ to $9$. 
The R-R sector has the axion field $\chi$, 
another anti-symmetric tensor field $B^{(2)}_{MN}$ and 
a rank four anti-symmetric tensor field $D_{MNPQ}$ whose field strength 
is self-dual. 
Although there is no action in ten dimensions giving us the full field 
equations, 
one can write down an action giving rise to the field equations if 
$D_{MNPQ}\equiv 0$ \cite{BHO}. 
Then we assume this condition. 
The action is 
\begin{eqnarray}
S_{IIB}=\int d^{10}x\sqrt{-\;^{10}\!g}\left( e^{\Phi}\left[-\;^{10}\!R
-(D\Phi)^{2}+\frac{1}{12}(H^{(1)})^{2}\right]
+\frac{1}{2}(D\chi)^{2}+\frac{1}{12}(H^{(1)}\chi+H^{(2)})^{2}\right),
\end{eqnarray}
where $\;^{10}\!R$ is the ten-dimensional Ricci scalar, 
$\;^{10}\!g:=\det{g_{MN}}$, $D$ is the covariant derivative of $g_{MN}$ 
and $H^{(i)}$ are the field strengths of 
the two-form fields $B^{(i)}$. 
This ten-dimensional effective action can be expressed 
in the manifestly self-S-duality covariant form 
in the Einstein frame \cite{H}.  
The toroidal compactification of the action to lower-dimensions, 
preserving self-S-duality, has been obtained \cite{MJ}. 
Thus, we will suppose such compactification. 
Furthermore, some restrictions are imposed for simplicity. 
In summarize, 
\\
\\
{\bf Assumptions}
{\it 
\begin{enumerate}
\item The simplest toroidal ansatz
$$
ds_{10}=\tilde{g}_{\mu\nu}(x)dx^{\mu}dx^{\nu}+e^{\beta(x)/\sqrt{3}}
\delta_{ab}dX^{a}dX^{b},
$$
where $\beta$ is a moduli field describing the volume of the internal 
space, $\mu$, $\nu$ run from $0$ to $3$ and $a$, $b$ run from $4$ to $9$.
\item All fields are independent of the internal coordinate $X^{a}$.
\item There are no vector or moduli fields coming from the compactification 
of the metric or the three-forms.
\hfill$\Box$
\end{enumerate}
}
\noindent
Under the above assumptions, 
the reduced four-dimensional dual action in the Einstein frame 
$$
g_{\mu\nu}=e^{-\phi}\tilde{g}_{\mu\nu}
$$
is given by \cite{LWC}
\begin{eqnarray}
\label{action}
S_{4}=\int d^{4}x\left[{\cal L}_{G}+{\cal L}_{M}\right]
\end{eqnarray}
\begin{eqnarray}
\label{actionG+M}
{\cal L}_{G}=-\sqrt{-g}\;^{4}\!R,
\end{eqnarray}
\begin{eqnarray}
{\cal L}_{M}=\frac{\sqrt{-g}}{2}\left[(\nabla\phi)^{2}+(\nabla\beta)^{2}
+\frac{1}{6}e^{-2\phi}(H^{(1)})^{2}
+e^{\sqrt{3}\beta+\phi}(\nabla\chi)^{2}
+\frac{1}{6}e^{\sqrt{3}\beta-\phi}(H^{(1)}\chi+H^{(2)})^{2}\right],
\end{eqnarray}
where $\;^{4}\!R$ is Ricci scalar with respect to four dimensional 
Lorentzian metric $g_{\mu\nu}$, $g=\det g_{\mu\nu}$, 
$\nabla$ is the covariant derivative with respect to $g_{\mu\nu}$ and 
$\phi:=\Phi -\sqrt{3}\beta$ 
is the effective dilaton field.

In four-dimensions, we can 
define two pseudo-scalar fields $\sigma_{i}$ 
dual to the NS-NS and R-R three-form field strength $H^{(i)}$, 
respectively.
\begin{eqnarray}
H^{(1)\mu\nu\lambda}=\epsilon^{\mu\nu\lambda\kappa}e^{2\phi}
(\nabla_{\kappa}\sigma_{1}-\chi\nabla_{\kappa}\sigma_{2}),
\end{eqnarray}
\begin{eqnarray}
H^{(2)\mu\nu\lambda}=\epsilon^{\mu\nu\lambda\kappa}
[e^{\phi-\sqrt{3}\beta}\nabla_{\kappa}\sigma_{2}
-\chi e^{2\phi}(\nabla_{\kappa}\sigma_{1}-\chi\nabla_{\kappa}\sigma_{2})].
\end{eqnarray}
In the dual formulation, the effective action for matter fields is as follows. 
\begin{eqnarray}
\label{actiondual}
{\cal L}_{M}=\frac{\sqrt{-g}}{2}\left[(\nabla\phi)^{2}+(\nabla\beta)^{2}
+e^{\sqrt{3}\beta+\phi}(\nabla\chi)^{2}
+e^{-\sqrt{3}\beta+\phi}(\nabla\sigma_{2})^{2}
+e^{2\phi}(\nabla\sigma_{1}-\chi\nabla\sigma_{2})^{2}\right].
\end{eqnarray}
As mentioned earlier, this action has self-S-dual symmetry 
(see appendix \ref{Self-S-duality}).

Varying the action~(\ref{action}) with respect to the functions, 
we have the following field equations.
\begin{eqnarray}
G_{\mu\nu}&=&T_{\mu\nu}\nonumber \\
&=&\frac{1}{2}
[\nabla_{\mu}\phi\nabla_{\nu}\phi+\nabla_{\mu}\beta\nabla_{\nu}\beta
+e^{\sqrt{3}\beta+\phi}\nabla_{\mu}\chi\nabla_{\nu}\chi
+e^{-\sqrt{3}\beta+\phi}\nabla_{\mu}\sigma_{2}\nabla_{\nu}\sigma_{2}
\nonumber \\
&&+e^{2\phi}(\nabla_{\mu}\sigma_{1}-\chi\nabla_{\mu}\sigma_{2})
(\nabla_{\nu}\sigma_{1}-\chi\nabla_{\nu}\sigma_{2})]
\nonumber \\
&&-\frac{1}{4}g_{\mu\nu}
[(\nabla\phi)^{2}+(\nabla\beta)^{2}+e^{\sqrt{3}\beta+\phi}(\nabla\chi)^{2}
+e^{-\sqrt{3}\beta+\phi}(\nabla\sigma_{2})^{2}
+e^{2\phi}(\nabla\sigma_{1}-\chi\nabla\sigma_{2})^{2}],
\end{eqnarray}
\begin{eqnarray}
\nabla_{\mu}\nabla^{\mu}\phi
-\frac{1}{2}e^{\sqrt{3}\beta+\phi}(\nabla\chi)^{2}
-\frac{1}{2}e^{-\sqrt{3}\beta+\phi}(\nabla\sigma_{2})^{2}
-e^{2\phi}
(\nabla\sigma_{1}-\chi\nabla\sigma_{2})^{2}=0 ,
\end{eqnarray}
\begin{eqnarray}
\nabla_{\mu}\nabla^{\mu}\beta
-\frac{\sqrt{3}}{2}e^{\sqrt{3}\beta+\phi}(\nabla\chi)^{2}
+\frac{\sqrt{3}}{2}e^{-\sqrt{3}\beta+\phi}(\nabla\sigma_{2})^{2}=0,
\end{eqnarray}
\begin{eqnarray}
\nabla_{\mu}\nabla^{\mu}\chi
+(\sqrt{3}\nabla_{\mu}\beta+\nabla_{\mu}\phi)\nabla^{\mu}\chi
+e^{-\sqrt{3}\beta+\phi}(\nabla^{\mu}\sigma_{1}-\chi\nabla^{\mu}\sigma_{2})
\nabla_{\mu}\sigma_{2}=0,
\end{eqnarray}
\begin{eqnarray}
\nabla_{\mu}\nabla^{\mu}\sigma_{1}&-&\nabla_{\mu}\chi\nabla^{\mu}\sigma_{2}
+2\nabla_{\mu}\phi(\nabla^{\mu}\sigma_{1}-\chi\nabla^{\mu}\sigma_{2})
\nonumber \\
&+&\chi[\nabla^{\mu}\sigma_{2}(-\sqrt{3}\nabla_{\mu}\beta+\nabla_{\mu}\phi)
-e^{\sqrt{3}\beta+\phi}\nabla_{\mu}\chi
(\nabla^{\mu}\sigma_{1}-\chi\nabla^{\mu}\sigma_{2})]=0,
\end{eqnarray}
\begin{eqnarray}
\nabla_{\mu}\nabla^{\mu}\sigma_{2}
+(-\sqrt{3}\nabla_{\mu}\beta+\nabla_{\mu}\phi)\nabla^{\mu}\sigma_{2}
-e^{\sqrt{3}\beta+\phi}[\nabla_{\mu}\chi
(\nabla^{\mu}\sigma_{1}-\chi\nabla^{\mu}\sigma_{2})]=0.
\end{eqnarray}

Now, let us fix spacetimes we consider as follows. 
\begin{definition}{\rm \cite{GR,CP}}
Spacetimes are {\rm $T^{3}$-Gowdy symmetric spacetimes} if the following 
conditions hold.
\begin{enumerate}
\item Compact, connected, orientable Cauchy surfaces exist.
\item There is an effective action of $U(1)\times U(1)$ 
isometry group on the Cauchy surfaces.
\item The twist constants associated to the $U(1)\times U(1)$ 
isometry group vanish.
\item Topology of the Cauchy surfaces is $T^{3}$.
\end{enumerate}
\end{definition}
A metric of Gowdy symmetric spacetimes can be written by 
\begin{eqnarray}
\label{metric}
ds=e^{\lambda/2}t^{-1/2}(-dt^{2}+d\theta^{2})
+R[e^{-Z}(dy+Xdz)^{2}+e^{Z}dz^{2}].
\end{eqnarray}
$\frac{\partial}{\partial y}$ and $\frac{\partial}{\partial z}$ 
are spacelike Killing vectors 
and each direction of space is periodic. 
Then, the metric functions $\lambda$, $R$, $Z$, $X$ and 
functions describing matter fields $\phi$, $\beta$, $\chi$, $\sigma_{i}$ 
depend on time coordinate 
$t$ and space coordinate $\theta$ and they are periodic in $\theta$.

Under the metric (\ref{metric}), 
we have the following Lagrangean density for type IIB superstring theory. 
\begin{eqnarray}
{\cal L}_{G}=\frac{1}{2R}\left[\dot{R}^{2}-R'^{2}+4R(\ddot{R}-R'')\right]
-\frac{R}{2}\left[\frac{1}{t^{2}}+\ddot{\lambda}-\lambda''+\dot{Z}^{2}-Z'^{2}
+e^{-2Z}(\dot{X}^{2}-X'^{2})\right],
\end{eqnarray}
and 
\begin{eqnarray}
{\cal L}_{M}=
-\frac{R}{2}
\lgroup
\dot{\phi}^{2}-\phi'^{2}+\dot{\beta}^{2}-\beta'^{2}
&+&e^{\sqrt{3}\beta+\phi}(\dot{\chi}^{2}-\chi'^{2})\nonumber \\
&+&e^{-\sqrt{3}\beta+\phi}(\dot{\sigma_{2}}^{2}-\sigma_{2}'^{2})
+e^{2\phi}\{(\dot{\sigma_{1}}-\chi\dot{\sigma_{2}})^{2}
-(\sigma_{1}'-\chi\sigma_{2}')^{2}\}
\rgroup,
\end{eqnarray}
where dot and prime denote partial derivative with respect to 
$t$ and $\theta$, respectively.

Now, we must fix {\it time}, which should be chosen geometrically. 
From the Einstein-matter equations, 
$G_{tt}-G_{\theta\theta}=T_{tt}-T_{\theta\theta}$, 
we have the following one.
\begin{eqnarray}
\ddot{R}-R''=0.
\end{eqnarray}
Then, we can take the {\it areal coordinate} globally \cite{GR,MV}, 
that is $R=t$, 
since we assume that topology of space is $T^{3}$. As one can see in the 
metric (\ref{metric}), geometrical meaning of 
$R$ is the area function of the orbit of the isometry group. 
Under this time coordinate, the Einstein-matter equations are as follows.

\noindent
\\
{\it Constraint equations} 
\begin{eqnarray}
\label{hamiltonianconstraint}
\dot{\lambda}=t[\dot{Z}^{2}+Z'^{2}+e^{-2Z}(\dot{X}^{2}+X'^{2})+4T_{tt}],
\end{eqnarray}
where
\begin{eqnarray}
4T_{tt}=\dot{\phi}^{2}+\phi'^{2}+\dot{\beta}^{2}+\beta'^{2}
&+&
e^{\sqrt{3}\beta+\phi}(\dot{\chi}^{2}+\chi'^{2})+
e^{-\sqrt{3}\beta+\phi}(\dot{\sigma_{2}}^{2}+\sigma_{2}'^{2})\nonumber \\
&+&
e^{2\phi}[(\dot{\sigma_{1}}-\chi\dot{\sigma_{2}})^{2}+
(\sigma_{1}'-\chi\sigma_{2}')^{2}],
\end{eqnarray}
and 
\begin{eqnarray}
\label{momentumconstraint}
\lambda'=2t[\dot{Z}Z'+e^{-2Z}\dot{X}X'+2T_{t\theta}],
\end{eqnarray}
where
\begin{eqnarray}
2T_{t\theta}=
\dot{\phi}\phi'+\dot{\beta}\beta'+e^{\sqrt{3}\beta+\phi}\dot{\chi}\chi'
+e^{-\sqrt{3}\beta+\phi}\dot{\sigma_{2}}\sigma_{2}'
+e^{2\phi}(\dot{\sigma_{1}}-\chi\dot{\sigma_{2}})(\sigma_{1}'-\chi\sigma_{2}')
.
\end{eqnarray}
\\
{\it Evolution equations}
\begin{eqnarray}
\label{lambda}
\ddot{\lambda}-\lambda''&+&\dot{Z}^{2}-Z'^{2}+e^{-2Z}(\dot{X}^{2}-X'^{2})
+\dot{\phi}^{2}-\phi'^{2}+\dot{\beta}^{2}-\beta'^{2}\nonumber \\
&+&e^{\sqrt{3}\beta+\phi}(\dot{\chi}^{2}-\chi'^{2})
+e^{-\sqrt{3}\beta+\phi}(\dot{\sigma_{2}}^{2}-\sigma_{2}'^{2})
+e^{2\phi}[(\dot{\sigma_{1}}-\chi\dot{\sigma_{2}})^{2}
-(\sigma_{1}'-\chi\sigma_{2}')^{2}]=0,
\end{eqnarray}
\begin{eqnarray}
\label{Z}
\Box_{G}Z=-e^{-2Z}(\dot{X}^{2}-X'^{2}),
\end{eqnarray}
\begin{eqnarray}
\label{X}
\Box_{G}X=2(\dot{Z}\dot{X}-Z'X'),
\end{eqnarray}
\begin{eqnarray}
\label{phi}
\Box_{G}\phi=\frac{1}{2}\left[
e^{\sqrt{3}\beta+\phi}(\dot{\chi}^{2}-\chi'^{2})
+e^{-\sqrt{3}\beta+\phi}(\dot{\sigma_{2}}^{2}-\sigma_{2}'^{2})
+2e^{2\phi}\{(\dot{\sigma_{1}}-\chi\dot{\sigma_{2}})^{2}
-(\sigma_{1}'-\chi\sigma_{2}')^{2}\}
\right],
\end{eqnarray}
\begin{eqnarray}
\label{beta}
\Box_{G}\beta=\frac{\sqrt{3}}{2}\left[
e^{\sqrt{3}\beta+\phi}(\dot{\chi}^{2}-\chi'^{2})
-e^{-\sqrt{3}\beta+\phi}(\dot{\sigma_{2}}^{2}-\sigma_{2}'^{2})
\right],
\end{eqnarray}
\begin{eqnarray}
\label{sigma1}
\Box_{G}\sigma_{1}&=&(\dot{\chi}\dot{\sigma_{2}}-\chi'\sigma_{2}')
-2\{(\dot{\sigma_{1}}-\chi\dot{\sigma_{2}})\dot{\phi}
-(\sigma_{1}'-\chi\sigma_{2}')\phi'\}\nonumber \\
&-&\chi\left[(-\sqrt{3}\dot{\beta}+\dot{\phi})\dot{\sigma_{2}}
-(-\sqrt{3}\beta'+\phi')\sigma_{2}'
-e^{\sqrt{3}\beta+\phi}\{(\dot{\sigma_{1}}-\chi\dot{\sigma_{2}})\dot{\chi}
-(\sigma_{1}'-\chi\sigma_{2}')\chi'\}\right],
\end{eqnarray}
\begin{eqnarray}
\label{sigma2}
\Box_{G}\sigma_{2}=
-(-\sqrt{3}\dot{\beta}+\dot{\phi})\dot{\sigma_{2}}
+(-\sqrt{3}\beta'+\phi')\sigma_{2}'
+e^{\sqrt{3}\beta+\phi}\left[(\dot{\sigma_{1}}-\chi\dot{\sigma_{2}})\dot{\chi}
-(\sigma_{1}'-\chi\sigma_{2}')\chi'\right],
\end{eqnarray}
\begin{eqnarray}
\label{chi}
\Box_{G}\chi=
-(\sqrt{3}\dot{\beta}+\dot{\phi})\dot{\chi}
+(\sqrt{3}\beta'+\phi')\chi'
-e^{-\sqrt{3}\beta+\phi}
\left[(\dot{\sigma_{1}}-\chi\dot{\sigma_{2}})\dot{\sigma_{2}}
-(\sigma_{1}'-\chi\sigma_{2}')\sigma_{2}'\right],
\end{eqnarray}
where $$\Box_{G}:=\frac{\partial^{2}}{\partial t^{2}}
+\frac{1}{t}\frac{\partial}{\partial t}
-\frac{\partial^{2}}{\partial \theta^{2}}.$$
We call the above system consisting of equations 
(\ref{hamiltonianconstraint})-(\ref{chi}) 
{\it $T^{3}$-Gowdy symmetric IIB system} and 
{\it $T^{3}$-Gowdy symmetric IIB spacetimes} is defined as 
solutions to $T^{3}$-Gowdy symmetric IIB system. 

Equation (\ref{lambda}) can be obtained from other equations. 
Thus, the metric function $\lambda$ is decoupled with other functions. 
Therefore, $\lambda$ is evaluated by the constraint equations 
(\ref{hamiltonianconstraint}) and (\ref{momentumconstraint}) after 
obtaining estimates of other functions.

\section{Global existence theorem in the areal time}

We now state a global existence theorem as follows. 
\begin{theorem}
\label{thm1}
Let $({\cal M}, g_{\mu\nu}, \phi, \beta, \chi, \sigma_{1}, \sigma_{2})$ be 
the maximal Cauchy development of $C^{\infty}$ initial data for the 
$T^{3}$-Gowdy symmetric IIB system. 
Then ${\cal M}$ can be covered by the areal coordinate with $t\in(0,\infty)$.
\end{theorem}

The system of evolution equations (\ref{Z})-(\ref{chi}) is one of semi-linear 
wave equations. 
Therefore, the local existence theorem is shown by standard arguments. 
In order to show that solution to $T^{3}$-Gowdy symmetric IIB system can be 
extended to $t\rightarrow +\infty$ all that needs to be done 
for all finite time.

\begin{definition}
Let $M$ and $N$ be manifolds 
with Lorentzian metric $\;^{B}\!g$ and Riemannian 
metric $\;^{T}\!g$, respectively. 
For a map $\Psi :M\rightarrow N$ we define the Lagrangian
\begin{eqnarray}
\label{wave-map-lag}
L[\Psi]=\frac{1}{2}\int_{M}\langle d\Psi, d\Psi\rangle
=\frac{1}{2}\int_{M}
\;^{B}\!g^{\alpha\beta}\;^{T}\!g_{ab}\partial_{\alpha}\Psi^{a}
\partial_{\beta}\Psi^{b}d\nu_{M}.
\end{eqnarray}
The resulting Euler-Lagrange equations will be hyperbolic 
and the solutions are called {\rm wave maps}. 
In local coordinates, the equation for $\Psi$ is 
\begin{eqnarray}
\label{wave-map-eq}
\Box_{B}\Psi^{a}+\Gamma^{a}_{bc}(\Psi)\partial_{\alpha}\Psi^{b}
\partial^{\alpha}\Psi^{c}=0,
\end{eqnarray}
where $\Box_{B}$ is the d'Alembertian with respect to $\;^{B}\!g$ and 
$\Gamma$ is the Christoffel symbol of the target manifold $N$.
\end{definition}
\begin{lemma}
\label{wave-map-l}
Let $M={\rm R}\times{\rm T}^{2}$ and $N={\rm R}^{7}$ be manifolds 
with Lorentzian metric $\;^{B}\!g$ and Riemannian 
metric $\;^{T}\!g$, respectively. 
The wave map $\Psi :M\rightarrow N$ is equivalent to the evolution 
equations (\ref{Z})-(\ref{chi}) of the $T^{3}$-Gowdy symmetric IIB system. 
Here,
\begin{eqnarray}
\;^{B}\!g=-dt^{2}+d\theta^{2}+t^{2}d\psi^{2},\hspace{.5cm}0\leq t, 
\hspace{.5cm}0\leq\theta,\psi\leq2\pi,
\end{eqnarray}
\begin{eqnarray}
\label{target}
\;^{T}\!g=\;^{T}\!g_{H}\oplus\;^{T}\!g_{HS},
\end{eqnarray}
where 
\begin{eqnarray}
\;^{T}\!g_{H}:=dZ^{2}+e^{-2Z}dX^{2},
\end{eqnarray}
\begin{eqnarray}
\;^{T}\!g_{HS}:=d\phi^{2}+d\beta^{2}
+e^{2\phi}(d\sigma_{1}-\chi d\sigma_{2})^{2}
+e^{\sqrt{3}\beta+\phi}d\chi^{2}+e^{-\sqrt{3}\beta+\phi}d\sigma_{2}^{2},
\end{eqnarray}
and $\Psi$ is independent of $\psi$.
\end{lemma}
{\it Proof of lemma \ref{wave-map-l}}: We can verify the above lemma by 
direct calculation. See appendix \ref{proof}.
\hfill$\Box$
\\
\\
{\it Remark}:
The target manifold is direct product of a hyperbolic space, whose metric 
is $\;^{T}\!g_{H}$, 
and a Riemannian space, whose metric is 
$\;^{T}\!g_{HS}$. 
If one consider vacuum Gowdy spacetimes, the former hyperbolic space 
is used.
The latter target space represents
the SL(3,R)/SO(3) coset corresponding to 
a homogeneous symmetric space \cite{LWC}.
Note that the action (\ref{actiondual}) is invariant under global SL(3,R) 
transformations. This symmetry causes the target space with the metric 
$\;^{T}\!g_{HS}$.   
\hfill$\Box$
\\

Thus, our problem has come to one of the wave map. 
Note that Shatah's elegant method for a wave map 
from a $(1+1)$-Minkowski spacetime to a Riemannian manifold 
\cite{SJ} 
can not be applied to our problem, 
since our wave map is not so, although it is a $(1+1)$ PDE system. 
For global existence and regularity of wave maps 
for smooth data in $(2+1)$-Minkowski 
spacetime to a Riemannian manifold, 
general results have not been known yet \cite{SS}. 
Here, a global existence theorem for $T^{3}$-Gowdy symmetric IIB spacetimes 
is shown by 
using light cone argument and a differential identity 
for the energy momentum tensor of wave maps.

The energy-momentum tensor ${\cal T}_{\alpha\beta}$ associated with the 
Lagrangian (\ref{wave-map-lag}) has the form
\begin{eqnarray}
{\cal T}_{\alpha\beta}=\;^{T}\!g_{ab}(\partial_{\alpha}\Psi^{a}
\partial_{\beta}\Psi^{b}
-\frac{1}{2}\;^{B}\!g_{\alpha\beta}\partial_{\alpha}\Psi^{a}
\partial^{\alpha}\Psi^{b}),
\end{eqnarray}
where the Greek indices denote $t$, $\theta$, $\psi$ and the Roman indices 
denote $Z$, $X$, $\phi$, $\beta$, $\sigma_{1}$, $\sigma_{2}$, $\chi$. 
For wave maps, ${\cal T}_{\alpha\beta}$ satisfied the local conservation laws 
\begin{eqnarray}
\label{conservation}
\;^{B}\!\nabla^{\alpha}{\cal T}_{\alpha\beta}=0,
\end{eqnarray}
where $\;^{B}\!\nabla$ is covariant differentiation with respect to the metric 
$\;^{B}\!g$. 

In our case, nonzero components of ${\cal T}_{\alpha\beta}$ are as follows, 
with $\langle ,\rangle$ and $\Vert\bullet\Vert$ denoting the inner product 
and the norm with respect to $\;^{T}\!g$.

\begin{eqnarray}
{\cal T}_{tt}=\frac{1}{2}(\Vert\dot{\Psi}\Vert^{2}+\Vert\Psi'\Vert^{2})
=:{\cal E},
\end{eqnarray}
\begin{eqnarray}
{\cal T}_{t\theta}=\langle\dot{\Psi},\Psi'\rangle
=:{\cal F},
\end{eqnarray}
\begin{eqnarray}
{\cal T}_{\theta\theta}
=\frac{1}{2}(\Vert\dot{\Psi}\Vert^{2}+\Vert\Psi'\Vert^{2})
={\cal T}_{tt},
\end{eqnarray}
and
\begin{eqnarray}
{\cal T}_{\psi\psi}=-\frac{t^{2}}{2}
(-\Vert\dot{\Psi}\Vert^{2}+\Vert\Psi'\Vert^{2})
=:-t^{2}{\cal G}.
\end{eqnarray}
Clearly, the following inequality holds: 
\begin{eqnarray}
\label{g<e}
\vert{\cal G}\vert\leq {\cal E}.
\end{eqnarray}

Now let $Y=f(t,\theta)\partial_{t}+g(t,\theta)\partial_{\theta}$ be a 
smooth vector field on the base manifold $M$. 
Using the conservation laws (\ref{conservation}) we obtain 
a differential identity \cite{CTZ1}
\begin{eqnarray}
\label{wmid}
\;^{B}\!\nabla_{\alpha}({\cal T}^{\alpha}_{\beta}Y^{\beta})
=\frac{1}{2}{\cal T}^{\alpha\beta}\pi_{\alpha\beta},
\end{eqnarray}
where $\pi_{\alpha\beta}={\cal L}_{Y}\;^{B}\!g_{\alpha\beta}$ and 
its nonzero components are as follows:
\begin{eqnarray}
\pi_{tt}=-2\dot{f},
\end{eqnarray}
\begin{eqnarray}
\pi_{t\theta}=-f'+\dot{g},
\end{eqnarray}
\begin{eqnarray}
\pi_{\theta\theta}=2g',
\end{eqnarray}
and 
\begin{eqnarray}
\pi_{\psi\psi}=2tf.
\end{eqnarray}
Taking $f=1$ and $g=0$ into the equation (\ref{wmid}), we have 
\begin{eqnarray}
\label{f1g0}
\partial_{t}(t{\cal E})-\partial_{\theta}(t{\cal F})={\cal G}.
\end{eqnarray}
Similarly, by taking $f=0$ and $g=1$ we obtain
\begin{eqnarray}
\label{f0g1}
\partial_{t}(t{\cal F})-\partial_{\theta}(t{\cal E})=0.
\end{eqnarray}
Now we can show the following lemma. 
\begin{lemma}
\label{energy-estimate}
There is a positive constant $C$ such that
\begin{equation}
\label{lemma2}
   {\cal E} \leq C (1 + 1/t^{2}), \quad t \in (0, \infty), \nonumber
\end{equation}
where $C$ depends only on the initial data at $t = t_0$.
\end{lemma}
{\it Remark}: 
This result holds for any Riemannian target manifolds since 
the following proof is independent of geometry of them. 
\hfill$\Box$\\

\noindent{\it Proof of lemma \ref{energy-estimate}}: 
Let us define
$$
\tilde{{\cal E}}:=t{\cal E}, \hspace{.5cm}\tilde{{\cal F}}:=t{\cal F}.
$$
From equations (\ref{f1g0}) and (\ref{f0g1})
\begin{eqnarray}
\label{e-f}
\partial_{\xi}(\tilde{{\cal E}}-\tilde{{\cal F}})
={\cal G},
\end{eqnarray}
and 
\begin{eqnarray}
\label{e+f}
\partial_{\eta}(\tilde{{\cal E}}+\tilde{{\cal F}})
={\cal G},
\end{eqnarray}
where $\partial_{\xi}:=\partial_{t}+\partial_{\theta}$ and 
$\partial_{\eta}:=\partial_{t}-\partial_{\theta}$.

Now, we will use light cone argument for our problem. 
Let us consider the future direction. 
Integrating equations (\ref{e-f}) and (\ref{e+f}) along null paths starting at 
$(\hat{t},\hat{\theta})$ and ending at the initial $t_{0}$-Cauchy surface, 
where $0<t_{0}\leq\hat{t}$, and adding these equations, we have 
\begin{eqnarray}
\label{add}
2\tilde{{\cal E}}(\hat{t},\hat{\theta})
=&&[\tilde{{\cal E}}(t_{0},\hat{\theta}+\hat{t}-t_{0})
+\tilde{{\cal E}}(t_{0},\hat{\theta}-\hat{t}+t_{0})
+\tilde{{\cal F}}(t_{0},\hat{\theta}-\hat{t}+t_{0})
-\tilde{{\cal F}}(t_{0},\hat{\theta}+\hat{t}-t_{0})]\nonumber \\
&+&\int^{\hat{t}}_{t_{0}}{\cal G}(s,\hat{\theta}+s-t_{0})
+{\cal G}(s,\hat{\theta}-s+t_{0})ds.
\end{eqnarray}
Taking supremums over all values of $\theta$ 
on the both side of equation (\ref{add}), we obtain 
\begin{eqnarray}
\sup_{\theta}\tilde{{\cal E}}(\hat{t},\theta)
\leq\sup_{\theta}\tilde{{\cal E}}(t_{0},\theta)
+\sup_{\theta}\tilde{{\cal F}}(t_{0},\theta)
+\int^{\hat{t}}_{t_{0}}\frac{1}{s}\sup_{\theta}\tilde{{\cal E}}(s,\theta)ds,
\end{eqnarray}
where inequality (\ref{g<e}) was used. 
From Gronwall's lemma \cite{SC} it follows that 
\begin{eqnarray}
\sup_{\theta}\tilde{{\cal E}}(\hat{t},\theta)
\leq[\sup_{\theta}\tilde{{\cal E}}(t_{0},\theta)
+\sup_{\theta}\tilde{{\cal F}}(t_{0},\theta)]
\exp\left[\int^{\hat{t}}_{t_{0}}\frac{1}{s}ds\right]
=[\sup_{\theta}\tilde{{\cal E}}(t_{0},\theta)
+\sup_{\theta}\tilde{{\cal F}}(t_{0},\theta)]\frac{\hat{t}}{t_{0}}.
\end{eqnarray}
Then,
\begin{eqnarray}
\sup_{\theta}{\cal E}(\hat{t},\theta)
\leq C_{1}, \hspace{.5cm}\forall \hat{t}\in[t_{0},\infty), 
\end{eqnarray}
where $C_{1}$ is a positive 
constant depending on only the initial data $t=t_{0}$.

We can use similar argument 
into the past direction. 
Indeed, integrating equations (\ref{e-f}) and (\ref{e+f}) 
along null paths starting at 
$(\hat{t},\hat{\theta})$ and ending at the initial $t_{0}$-Cauchy surface 
and repeating the previous argument, we have 
\begin{eqnarray}
\sup_{\theta}\tilde{{\cal E}}(\hat{t},\theta)
\leq[\sup_{\theta}\tilde{{\cal E}}(t_{0},\theta)
+\sup_{\theta}\tilde{{\cal F}}(t_{0},\theta)]
\exp\left[\int_{\hat{t}}^{t_{0}}\frac{1}{s}ds\right]
=[\sup_{\theta}\tilde{{\cal E}}(t_{0},\theta)
+\sup_{\theta}\tilde{{\cal F}}(t_{0},\theta)]\frac{t_{0}}{\hat{t}},
\end{eqnarray}
where $\hat{t}\in (0,t_{0}]$. Then, 
\begin{eqnarray}
\sup_{\theta}{\cal E}(\hat{t},\theta)
\leq \frac{C_{2}}{\hat{t}^{2}}, \hspace{.5cm}\forall \hat{t}\in(0,t_{0}]. 
\end{eqnarray}
Thus, we have the estimate (\ref{lemma2}) and ${\cal E}$ is bounded 
on every compact interval of $(0, \infty)$. 
\hfill$\Box$
\\
\\
\noindent
{\it Proof of theorem \ref{thm1}}:
From lemma \ref{energy-estimate}, we have the desired bounds on 
$\mid\dot{Z}\mid$, $\mid Z'\mid$, $\mid e^{-Z}\dot{X}\mid$, 
$\mid e^{-Z}X'\mid$, 
$\mid\dot{\phi}\mid$, $\mid \phi '\mid$, 
$\mid\dot{\beta}\mid$, $\mid \beta '\mid$, 
$\mid e^{\frac{\sqrt{3}\beta+\phi}{2}}\dot{\chi}\mid$, 
$\mid e^{\frac{\sqrt{3}\beta+\phi}{2}}\chi'\mid$, 
$\mid e^{\frac{-\sqrt{3}\beta+\phi}{2}}\dot{\sigma_{2}}\mid$, 
$\mid e^{\frac{-\sqrt{3}\beta+\phi}{2}}\sigma_{2}'\mid$, 
$\mid e^{\phi}(\dot{\sigma_{1}}-\chi\dot{\sigma_{2}})\mid$, 
$\mid e^{\phi}(\sigma_{1}'-\chi\sigma_{2}')\mid$ 
for all $t\in (0,\infty)$.
Once we have bounds on the first derivatives of $Z$, $\phi$ and $\beta$, 
it follows that $Z$, $\phi$ and $\beta$ are bound for all $t\in (0,\infty)$. 
Then, we have bounds on $\dot{X}$, $X'$, $\dot{\chi}$, $\chi'$, 
$\dot{\sigma_{2}}$, $\sigma_{2}'$, $\dot{\sigma_{1}}-\chi\dot{\sigma_{2}}$ 
and $\sigma_{1}'-\chi\sigma_{2}'$. 
Consequently, $X$, $\chi$, $\sigma_{2}$, $\dot{\sigma_{1}}$ and $\sigma_{1}'$ 
are bounded. 
Finally, we have boundedness on $\sigma_{1}$.

Next, we must show bounds on the second derivatives of the functions.  
There is a well-known general fact that, in order to ensure the continued 
existence of a solution of a system of semi-linear wave equations, 
it is enough to bound the first derivative pointwise.  
Then, we have boundedness of the higher derivatives.

By the constraint equations~(\ref{hamiltonianconstraint}) 
and (\ref{momentumconstraint}), 
boundedness for the function $\lambda$ is 
also shown. 
Indeed, we have the following equations:
\begin{eqnarray}
\label{constraints}
\dot{\lambda}=2t{\cal E} \hspace{1cm} {\rm and} 
\hspace{1cm}\lambda'=2t{\cal F}.
\end{eqnarray}
Since ${\cal E}$ and ${\cal F}$ are bounded as we have already seen, 
the first derivatives of $\lambda$ 
must be bounded uniformly for all $0<t<\infty$. 
Consequently, $\lambda$ itself also is bounded. 
Concerning the second derivative of $\lambda$, 
we can apply the previous argument again. 
Then, we have uniform $C^{2}$ bounds on all of the functions 
of $T^{3}$-Gowdy symmetric IIB system for all 
$t\in (0,\infty)$.

Finally, we must demand that the function $\lambda$ is compatible with 
the periodicity in $\theta$. 
This is true if $\lambda(t,-\pi)=\lambda(t,\pi)$ 
over the interval of existence. 
Integrating equation~(\ref{momentumconstraint}) 
for the interval $\theta\in[-\pi,\pi]$, 
we have a constraint
\begin{eqnarray}
\int^{\pi}_{-\pi}d\theta\lambda'=2\int^{\pi}_{-\pi}d\theta t{\cal F}=0.
\end{eqnarray}
This constraint condition need only be imposed on the initial Cauchy surface 
since this integral is conservation on any time interval if all other 
functions satisfy the periodicity condition. 
This fact follows from the constraint 
equations~(\ref{constraints}):
\begin{eqnarray}
\partial_{t}(t{\cal F})
=
\partial_{\theta}
(t{\cal E}),
\end{eqnarray}
where ${\cal E}$ consists of periodic functions. 
Thus, we have completed the proof of Theorem~\ref{thm1}.
\hfill$\Box$
\section{Global existence theorem in the CMC time}

It is known that global existence of 
CMC foliations would be shown for some Gowdy symmetric spacetimes with 
appropriate matter if one can get global foliations in the areal time 
for such spacetimes \cite{ARR}. 
Indeed, we can show the existence of global CMC foliations for 
$T^{3}$-Gowdy symmetric IIB spacetimes. 
An essential point of the arguments 
is the existence of a crushing singularity.

\begin{definition}
The spacetime singularity approached as $t$ tends to zero 
is called a {\it crushing singularity} if the mean curvature of spacelike 
hypersurfaces blows up uniformly as $t\rightarrow 0$. 
\end{definition}
\begin{proposition}
\label{crush}
$T^{3}$-Gowdy symmetric IIB spacetimes have  
crushing singularities. 
\end{proposition}
{\it Proof of proposition \ref{crush}}:
For the Gowdy metric (\ref{metric}) we can compute that
\begin{eqnarray}
\label{meancurvature}
tr K(t)
=-e^{-\frac{\lambda}{4}}t^{\frac{1}{4}}
\left[\frac{1}{4}\dot{\lambda}+\frac{3}{4t}\right],
\end{eqnarray}
where $tr K(t)$ is the mean curvature of a Cauchy hypersurface 
$\Sigma_{t}$ at the areal time $t$. 
The Hamiltonian constraint equation 
(\ref{hamiltonianconstraint}) is $\dot{\lambda}=2\tilde{{\cal E}}$. 
Then, the equation (\ref{meancurvature}) is 
\begin{eqnarray}
\label{meancurvature'}
tr K(t)
=-e^{-\frac{\lambda}{4}}t^{\frac{1}{4}}
{\cal B}_{t},
\end{eqnarray}
where ${\cal B}_{t}:=\frac{3}{4t}+\frac{1}{2}\tilde{{\cal E}}$. 
Note that 
\begin{eqnarray}
\label{bt}
\inf_{\Sigma_{t}}{\cal B}_{t}\geq\frac{3}{4t}.
\end{eqnarray}
Let us assume $0<t\leq t_{0}$. 
From the Hamiltonian constraint equation 
(\ref{hamiltonianconstraint}) 
\begin{eqnarray}
\int^{t_{0}}_{t}\dot{\lambda}ds=2\int^{t_{0}}_{t}\tilde{{\cal E}}ds.
\end{eqnarray}
Then, we have 
\begin{eqnarray}
\exp\left[{-\frac{\lambda(t)}{4}}\right]
=\exp\left[{-\frac{\lambda(t_{0})}{4}
+\frac{1}{2}\int^{t_{0}}_{t}\tilde{{\cal E}}ds}\right]
\geq\exp\left[{-\frac{\lambda(t_{0})}{4}}\right].
\end{eqnarray}
Thus, we obtain the following inequality for $t\in (0,t_{0}]$:
\begin{eqnarray}
\label{mestimate}
tr K(t)\leq -C_{3}t^{\frac{1}{4}}{\cal B}_{t}, 
\end{eqnarray}
where $C_{3}>0$ is a constant. 
From equations (\ref{bt}) and (\ref{mestimate}) it is shown 
that $T^{3}$-Gowdy symmetric IIB spacetime 
has a crushing singularity into the past 
because of 
\begin{eqnarray}
\sup_{\Sigma_{t}}tr K(t)\leq -C_{4}t^{\frac{1}{4}}t^{-1}=-C_{4}t^{-\frac{3}{4}}
\rightarrow -\infty\hspace{.5cm}{\rm as}\hspace{.5cm}t\rightarrow 0.
\end{eqnarray}
\hfill$\Box$

\begin{theorem}
\label{cmc}
Let $({\cal M}, g_{\mu\nu}, \phi, \beta, \chi, \sigma_{1}, \sigma_{2})$ be 
the maximal Cauchy development of $C^{\infty}$ initial data for the 
$T^{3}$-Gowdy symmetric IIB system. 
Then ${\cal M}$ can be covered by the constant mean curvature hypersurfaces 
with $\tau\in(-\infty,0)$.
\end{theorem}
\noindent
{\it Sketch of the proof of theorem \ref{cmc}}: 
It was shown by Gerhardt \cite{GC} that a neighborhood of a crushing 
singularity can be foliated by CMC hypersurfaces. 
As we have already seen by proposition \ref{crush}, since the spacetime 
has a crushing singularity as the initial singularity, 
there is a CMC foliation in $D^{-}(\Sigma_{t_{0}})$. 
Once it is shown that the existence of a CMC hypersurface, 
say $\Sigma_{\tau_{0}}$,  
one can prove that $D^{+}(\Sigma_{\tau_{0}})$ 
admits a unique, monotonic CMC foliation which covers 
$D^{+}(\Sigma_{\tau_{0}})$ by Lemma 2 of the previous paper of the 
author \cite{N}
\footnote{In the lemma, the EMDA 
system is considered. As getting a global existence theorem in the areal 
coordinate, the same result can be also obtained in IIB system since 
non-negativity of $T_{tt}$ is used as a property of matter fields.}. 
Thus, the global existence result of CMC foliations is obtained 
and CMC time $\tau$ takes all values in the range $(-\infty,0)$. 
\hfill$\Box$
\section{Comments}
We have not discussed on the second step to prove the SCC, that is, 
inextendibility of maximal Cauchy development. 
We would like to comment on it. 
Into the future direction, 
we must show future completeness of any causal geodesic. 
The light cone argument and the CTZ's identity used to prove 
our theorems have many applicabilities. 
The tools would be able to use 
for investigations of asymptotic behavior into the future of 
the spacetimes \cite{CTZ2}.
Into the past direction, 
we have to prove blowup of the Kretschemann invariant 
${\cal K}:=R^{\mu\nu\lambda\sigma}R_{\mu\nu\lambda\sigma}$ 
as $t\rightarrow 0$. Unfortunately, 
there is no idea to estimate ${\cal K}$ at the present time. 
After solving these problem, 
we can conclude about inextendibility and then, would prove the (in)validity 
of the SCC conjecture for Gowdy symmetric IIB spacetimes.

\begin{center}
{\bf Acknowledgments}
\end{center}
I am grateful to Alan Rendall and 
Yoshio Tsutsumi for commenting on the manuscript. 

\appendix
\section{Self-S-duality}
\label{Self-S-duality}
Put $u:=\frac{\phi}{2}+\frac{\sqrt{3}\beta}{2}(=\frac{\Phi}{2})$ 
and $v:=\frac{\sqrt{3}\phi}{2}-\frac{\beta}{2}$. 
The action (\ref{actionG+M}) takes the form 
\begin{eqnarray}
\label{actionmodify}
S_{4}=\int d^{4}x\sqrt{-g}\left(-\;^{4}\!R+
\frac{1}{2}\left[(\nabla u)^{2}
+e^{2u}(\nabla\chi)^{2}+(\nabla v)^{2}
+e^{\sqrt{3}v}\left[e^{-u}(\nabla\sigma_{2})^{2}
+e^{u}(\nabla\sigma_{1}-\chi\nabla\sigma_{2})^{2}\right]\right]\right).
\end{eqnarray}
Now taking 
\begin{eqnarray}
w:=\chi+ie^{-u},
\end{eqnarray}
\begin{eqnarray}
{\cal A}_{ij}:=\frac{1}{{\rm Im}\hspace{.1cm} w}
\left(\begin{array}{cc}
|w|^{2} & -{\rm Re}\hspace{.1cm} w \\
-{\rm Re}\hspace{.1cm} w & 1 \\
\end{array}\right),
\end{eqnarray}
\begin{eqnarray}
{\cal B}^{i}:=
\left(\begin{array}{c}
\nabla\sigma_{2} \\
\nabla\sigma_{1} \\
\end{array}\right).
\end{eqnarray}
Then, the action (\ref{actionmodify}) is 
\begin{eqnarray}
S_{4}=\int d^{4}x\sqrt{-g}\left(-\;^{4}\!R+\frac{1}{2}\left[
\frac{|\nabla w|^{2}}{({\rm Im}\hspace{.1cm} w)^{2}}+e^{\sqrt{3}v}
{\cal B}^{i}{\cal A}_{ij}{\cal B}^{j}+(\nabla v)^{2}\right]\right).
\end{eqnarray}
This action is invariant under the following SL(2,R) transformation:
\begin{eqnarray}
\bar{w}=\frac{aw+b}{cw+d},\hspace{.5cm} 
\bar{g}_{\mu\nu}=g_{\mu\nu},\hspace{.5cm}
\bar{v}=v,\hspace{.5cm}
\bar{{\cal B}}^{i}=\Lambda^{i}_{j}{\cal B}^{j},\hspace{.5cm}
\Lambda^{i}_{j}=
\left(\begin{array}{cc}
d & c\\
b & a \\
\end{array}\right),
\hspace{.5cm}
ad-bc=1.
\end{eqnarray}
Indeed, we can show 
\begin{eqnarray}
\nabla\bar{w}=\frac{\nabla w}{(cw+d)^{2}},\hspace{.5cm} 
{\rm Im}\hspace{.1cm}\bar{w}
=\frac{{\rm Im}\hspace{.1cm}w}{|cw+d|^{2}},\hspace{.5cm}
{\rm Re}\hspace{.1cm}\bar{w}
=\frac{ac|w|^{2}+(ad+bc){\rm Re}\hspace{.1cm}w+bd}{|cw+d|^{2}},
\end{eqnarray}
and 
\begin{eqnarray}
{\cal A}(\bar{w})=(\Lambda^{-1})^{T}{\cal A}(w)\Lambda^{-1}.
\end{eqnarray}
In particular, when $\chi=0$ for $a=d=0$ and $b=-c=-1$, we have 
\begin{eqnarray}
\bar{\Phi}=-\Phi,\hspace{.5cm}
\bar{g}_{\mu\nu}=g_{\mu\nu},\hspace{.5cm}
\bar{v}=v,\hspace{.5cm}
\bar{\sigma}_{2}=\sigma_{1},\hspace{.5cm}
\bar{\sigma}_{1}=-\sigma_{2}.
\end{eqnarray}
As $e^{\Phi}$ implies the string coupling, the above transformation means that 
the strong coupling regime of the four-dimensional IIB theory is mapped 
onto the weak coupling one and vice versa, that is, {\it self-S-duality}. 
\section{Proof of lemma \ref{wave-map-l}}
\label{proof}
It is easy to verify $\Box_{B}=\Box_{G}$. 
Nonzero components of the Christoffel symbol $\Gamma$ of $N$ are as follows.
$$
\Gamma^{Z}_{XX}=e^{2Z}, \hspace{.5cm}\Gamma^{X}_{ZX}=-1
=-\Gamma^{\sigma_{1}}_{\sigma_{1}\phi}, 
\hspace{.5cm}\Gamma^{\phi}_{\sigma_{1}\sigma_{1}}=-e^{2\phi},
\hspace{.5cm}\Gamma^{\phi}_{\sigma_{1}\sigma_{2}}=\chi e^{2\phi},
\hspace{.5cm}
\Gamma^{\phi}_{\chi\chi}=-\frac{1}{2}e^{\sqrt{3}\beta+\phi}
=\Gamma^{\sigma_{2}}_{\sigma_{1}\chi},
$$
$$
\Gamma^{\phi}_{\sigma_{2}\sigma_{2}}
=-\frac{1}{2}(2\chi^{2}e^{2\phi}+e^{-\sqrt{3}\beta+\phi}),
\hspace{.5cm}\Gamma^{\beta}_{\sigma_{2}\sigma_{2}}
=\frac{\sqrt{3}}{2}e^{-\sqrt{3}\beta+\phi},
\hspace{.5cm}
\Gamma^{\beta}_{\chi\chi}
=-\frac{\sqrt{3}}{2}e^{\sqrt{3}\beta+\phi},
\hspace{.5cm}\Gamma^{\sigma_{1}}_{\sigma_{1}\chi}
=-\frac{1}{2}e^{\sqrt{3}\beta+\phi},
$$
$$
\Gamma^{\sigma_{1}}_{\sigma_{2}\phi}
=-\frac{1}{2}\chi,
\hspace{.5cm}\Gamma^{\sigma_{1}}_{\sigma_{2}\beta}
=-\frac{\sqrt{3}}{2}\chi,
\hspace{.5cm}
\Gamma^{\sigma_{1}}_{\sigma_{2}\chi}
=\frac{1}{2}(\chi^{2}e^{\sqrt{3}\beta+\phi}-1),
\hspace{.5cm}\Gamma^{\sigma_{2}}_{\sigma_{2}\phi}
=\frac{1}{2}=\Gamma^{\chi}_{\chi\phi},
$$
$$
\hspace{.5cm}\Gamma^{\sigma_{2}}_{\sigma_{2}\beta}
=-\frac{\sqrt{3}}{2}=-\Gamma^{\chi}_{\chi\beta},
\hspace{.5cm}\Gamma^{\sigma_{2}}_{\sigma_{2}\chi}
=\frac{1}{2}\chi e^{\sqrt{3}\beta+\phi},
\hspace{.5cm}\Gamma^{\chi}_{\sigma_{1}\sigma_{2}}
=\frac{1}{2}e^{-\sqrt{3}\beta+\phi},
\hspace{.5cm}\Gamma^{\chi}_{\sigma_{2}\sigma_{2}}
=-\chi e^{-\sqrt{3}\beta+\phi}.
$$
From these, we can get the fact that equation (\ref{wave-map-eq}) under 
conditions of lemma \ref{wave-map-l} is equivalent with 
the system of PDEs (\ref{Z})-(\ref{chi}).
\hfill$\Box$

\end{document}